\documentclass[11pt]{article}%
\usepackage[iso-8859-7]{inputenc}
\usepackage{epsfig}
\usepackage{graphicx}
\usepackage{amsfonts}
\usepackage{amssymb}
\usepackage{amsthm}
\usepackage{amsmath}
\usepackage{multirow}
\usepackage{cite}
\usepackage{mathtools}%
\newcommand{\newc}{\newcommand}
\newc{\ra}{\rightarrow}
\newc{\lra}{\leftrightarrow}
\newc{\ov}{\overline}
\newc{\pa}{\partial}
\newc{\be}{\begin{equation}}
\newc{\ee}{\end{equation}}
\newc{\ba}{\begin{eqnarray}}
\newc{\ea}{\end{eqnarray}}
\newc{\n}{\nu}
\newc{\D}{\Delta}
\newc{\eps}{\epsilon}
\newc{\la}{\lambda}
\newc{\e}{\epsilon}
\newc{\nn}{\nonumber}

\begin{document}

\thispagestyle{empty}

\vskip 2truecm
\vspace*{3cm}

\begin{center}
{\large{\bf
 Knitting neutrino mass textures with or without Tri-Bi maximal mixing}}

\vspace*{1cm}
{\bf G.K. Leontaris$^{\,1}$, N.D. Vlachos$^{\,2}$}\\[3.5mm]
$^1\,$Theoretical Physics Division, Ioannina University, GR-45110 Ioannina, Greece \\[2.5mm]
$^2\,$Theoretical Physics
Division, Aristotle University, GR-54124 Thessaloniki, Greece

\end{center}
\vspace*{1cm}

\begin{center}
\textbf{Abstract}
\end{center}

The solar and baseline neutrino oscillation  data suggest bimaximal neutrino
mixing among the first two generations, and trimaximal mixing between all three
neutrino flavors. It has been conjectured that this indicates the existence of
an underlying symmetry for the leptonic fermion mass textures. The experimentally
measured quantities however, are associated to the latter indirectly and in a rather
complicated way through the mixing matrices of the charged leptons and neutrinos.
Motivated by these facts, we derive exact analytical expressions which directly link
the charged lepton and neutrino mass and mixing parameters to measured quantities and
obtain constraints on the parameter space. We discuss deviations from Tri-Bi mixing
matrices and present minimal extensions of the Harrison Perkins and Scott matrices
capable of interpreting all neutrino data.
\noindent

\vfill \newpage

\section{Introduction}

Since the experimental confirmation of   neutrino oscillations,
 there have been assiduous efforts  to   measure the exact mixing angles and their tiny
masses \footnote{For reviews see~\cite{Mohapatra:2005wg}-\cite{Altarelli:2010gt}.}.
The last few years we are in a position to know up
to a high accuracy the three neutrino mixing angles and the mass squared
differences. Several phenomenological explorations
have led to the conclusion that the so called Tri-Bi (TB) maximal
mixing~\cite{Harrison:2002er} is in remarkable agreement with the solar and
atmospheric neutrino data. Indeed, the experimental range of the three angles
lie in the range
\begin{align}
\sin^{2}\theta_{12}\approx0.312^{+0.019}_{-0.018} ,\;\sin^{2}\theta
_{23}\approx0.466^{+0.073}_{-0.058},\; \sin^{2}\theta_{13}\approx
0.126^{+0.053}_{-0.049}%
\end{align}
while the values of TB-prediction are $\frac13,\frac12, 0$ respectively.

Another interesting aspect is the fact that in the basis where the charged
lepton mass matrix is diagonal, TB-mixing is independent of the neutrino mass
eigenvalues, the symmetric neutrino mass elements need only to satisfy three
simple relations~\cite{Abbas:2010jw}
\[
m_{e\mu}=m_{e\tau},\;m_{\mu\mu}=m_{\tau\tau},\;m_{ee}+m_{e\mu}=m_{\mu\mu
}+m_{\mu\tau}\,\cdot
\]

One may attempt to attribute the regularity of the data in the leptonic sector
to the existence of some particular symmetry of  a suitable  theoretical model.
However, the above picture is definitely different from the corresponding one
in the quark sector thus, in view of  accumulating experimental evidence during the
last decade, the reconciliation of neutrino data with simple $U(1)$ family
symmetry models~\cite{Leontaris:1999wf} is rather unlikely. It has been
suggested for example, that this specific structure might originate from a
discrete non-Abelian symmetry~\cite{Lam:2008sh,deMedeirosVarzielas:2005qg}. A
different point of view is taken however in~\cite{Abbas:2010jw} where the
authors claim that TB-mixing might be completely accidental as they found that
significant violations from TB-mixing may occur within the present
experimental bounds, with $1-3$ mixing in particular leading to substantial
deviations. Several other suggestions including the introduction of discrete
and unified theories have appeared in the
literature~\cite{Xing:2002sw}-\cite{deMedeirosVarzielas:2006fc}.

The modifications on the TB mixing suggested by several of these proposals are
based on perturbative considerations of the original TB-mass textures and the
mixing matrix. However, in order to consistently study the effects on the mixing
and the tiny mass differences in the neutrino sector, a rather accurate approach is
required.  Further, a major issue the present days is the exact measurement of
the $\theta_{13}$ angle which in the original TB-model was assumed to be exactly zero,
while  data do allow for a small deviation. In addition, several measurable
effects depend crucially on the value of the $\theta_{13}$ angle. For example, the survival
 probability of the reactor neutrinos involves both $\theta_{13}$ and
 $\Delta m_{31}^2$~\cite{Bilenky:2001jq}, and current bounds allow a
 small value for $\theta_{13}$.  If this angle is non-zero indeed, its value is  sensitive to
 any modification of the mass matrix; in this case, approximate results and perturbative
 expansions may not be adequate to reliably determine  the
 measured parameters in terms of   mass textures  eventually dictated
  by some symmetry.
 The aim of this letter is to fill this gap.  Assuming only general hermitean
 mass squared matrices for the charge leptons and neutrinos,
 we will derive  analytical results  for the mixing and mass-squared differences.
  To this end, using well known theorems from the spectral theory of matrices,
  we  express a general
$3\times3$ hermitean fermion mass matrix as a second degree polynomial of a
unitary matrix. This way, we are able to disentangle the mass eigenstates which
appear only in the coefficients of this expansion. We  then express
the neutrino mixing angles as functions of variables that parametrize the unitary
matrices which diagonalize the charged lepton and  neutrino mass
matrices \footnote{We note that the method developed here is general and can
be applied equally well to the quark sector.} respectively. This procedure gives enough
flexibility to determine the experimentally allowed range of the parameters,  and,
at the same time seek for mass textures  eventually dictated by
some underlying symmetry. As an immediate benefit of this   approach we get
 a non-zero $\theta_{13}$  angle emerging even in the
 minimal TB scenario, provided that at least one non-zero phase in the neutrino texture
 is assumed. In addition,  we will present a second example where  compatible neutrino textures arise,
otherwise not accessible by perturbative treatment around the TB solution.

\section{Formulation}

\bigskip

As explained in the introduction, the TB-maximal mixing is compatible with
the observed neutrino data. However, the very specific form of the mass
matrices postulated in this approach can only be embedded in particular classes of
unified theories and even less string derived models. Early and present endeavors in
this direction for example involve the  rather promising $A_4$ symmetry
as far as the neutrino sector is concerned. Thus, $A_4$ can be
generated by  elements $S,T$ satisfying $S^2=T^3=(ST)^3=1$ and these
can be viewed~\cite{Altarelli:2005yx} as a subgroup of the modular group which plays
a fundamental role in string theory.  Nevertheless, a straightforward application
  to the quark sector  is not very satisfactory
 since it predicts unacceptably  small quark mixing,  while only   contrived
  variants can possibly reconcile data from both the
 quark and lepton sectors~\cite{Ma:2001dn,He:2006dk,Feruglio:2007uu,Bazzocchi:2008sp,Honda:2008rs}.
Other attempts to generate TB-mixing relying  on the non-abelian family symmetry $\Delta(27)$
give results which are also compatible with quark mixing and can in principle
be embedded in a unified gauge theory\cite{deMedeirosVarzielas:2006fc}.
However, the  usefulness of  parametrizations dictated by such symmetries
is limited within the prescribed scenario  and might not capture cases
exhibiting  other possible interesting properties beyond TB-mixing.
 Moreover, if we  seriously wish to exploit the idea that some other
 underlying symmetry is found hidden behind the regularity of the
neutrino data, pertubative investigations around the TB solution are highly
unlikely to have a chance. We should not also ignore  the variety of unified
or string models where the existing vacua along flat
 directions usually break symmetries in a hard way and lead to complicated  mass textures.
Renormalization group effects, as well as
instanton contributions~\cite{Blumenhagen:2006xt,Ibanez:2006da,Leontaris:2009ci,Cvetic:2010mm}
may further obscure the original symmetry.

Taking into account the above considerations, we infer that the complicated
picture of the model building landscape as well as the sensitivity of
the neutrino data on TB departures, call for a detailed and {\it exact}
treatment of the neutrino sector. To accomplish this,
in this section, we will develop a new formalism for describing  $3\times3$
mass matrices and their corresponding diagonalizing transformations.
In doing this, we will generalize a formalism which appears to be a
special property  of the  original TB construction, namely the independence
of the  mixing angles from the eigenvalues. This will be a built-in property of
our suggested formalism and  will facilitate the analysis of  the
complicated structure of the leptonic sector.

We  wish to analyze  general models based on GUTs, SUSY-GUTs and strings
 which predict a variety of fermion mass
textures $m_{f}$  not necessarily symmetric or Hermitian. In the
present analysis we will consider  the Hermitian squares $m_{f}%
m_{f}^{\dagger}$ of the $3\times3$ fermion mass matrices which capture the
physical properties of a whole class of fermion mass textures $m_{f}$.

A general Hermitean $3\times3$ matrix contains $9$ independent elements and
can be written as
\begin{equation}
H=i\ln U
\end{equation}
where $U$ a unitary matrix. Using the Caley-Hamilton theorem we can write
\begin{equation}
H=b_{1}I+b_{2}U+b_{3}U^{2}
\end{equation}
where $b_{1}$,$b_{2}$,$b_{3}$ are complex in general. Reversing the argument,
we propose to  write a Hermitean mass matrix $M$ in the form
\begin{equation}
M=b_{1}I+b_{2}U+b_{3}U^{2}\label{ExMU}%
\end{equation}
where $U$ is a unitary matrix and, without loss of generality, we assume that
$\det U=1$. Now, the standard CKM form for a unitary matrix contains four
independent elements and has determinant one. Adding six degrees of freedom (d.o.f.) from the
complex $b_i$ coefficients we have a total of ten, so one d.o.f. is redundant, and can be removed by
requiring one eigenvalue of $U$ to be one as described in the next paragraph.

Since $U$ and $M$ obviously commute, the above expression can be diagonalized
by means of a similarity transformation. Thus, once we have expressed a given
$M$ in terms of $U$, we can find its diagonalizing matrix simply by
diagonalizing $U$. A diagonal unitary matrix is uniquely defined by
\begin{equation}
U_{d}=%
\begin{bmatrix}
e^{ia_{1}} & 0 & 0\\
0 & e^{ia_{2}} & 0\\
0 & 0 & e^{ia_{3}}%
\end{bmatrix}
\ \cdot
\end{equation}
One phase can be absorbed into a redefinition of the coefficients $b_{2}$ and
$b_{3}$ and  taking into account the determinant condition, we end up with
\begin{equation}
U_{d}=%
\begin{bmatrix}
e^{ia} & 0 & 0\\
0 & 1 & 0\\
0 & 0 & e^{-ia}%
\end{bmatrix}
\
\end{equation}
where the ordering of the diagonal elements may vary. Thus, our unitary matrix
can always be chosen to  have one eigenvalue equal to one. Denoting by $m_{1}%
,m_{2},m_{3}$ the (real) eigenvalues of $M$, we have the equations

\begin{align}
m_{1} &  =b_{1}+b_{2}e^{i\alpha}+b_{3}e^{2i\alpha}\nonumber\\
m_{2} &  =b_{1}+b_{2}+b_{3}\label{b_m}\\
m_{3} &  =b_{1}+b_{2}e^{-i\alpha}+b_{3}e^{-2i\alpha}\nonumber \cdot
\end{align}
The solution of the above
system for $b_{1},b_{2},b_{3}$ gives
\ba
b_{1}&=&-\frac{1}{4}\csc^{2}\frac{\alpha}{2}\left(\frac{e^{ -\frac{3}{2}i\alpha}~m_{1}+e^{ \frac{3}{2}i\alpha}~m_{3}}{2\cos\frac a2}-~m_{2}\right)\label{b1}
\\
b_{2}&=&+\frac{1}{4}\csc^{2}\frac{\alpha}{2}\left(e^{ -i\alpha}  (m_{1}-m_{2})-e^{
i\alpha} \left(  m_{2}-m_{3}\right)  \right)\label{b2}
\\
b_{3}&=&-\frac{1}{4}\csc^2\frac{\alpha}{2}\frac{e^{ -\frac{i}{2}\alpha}( m_{1}%
-m_{2})-e^{ \frac{i}{2}\alpha} \left(  m_{2}-m_{3}\right)}{2\cos\frac a2}
\cdot\label{b3}
\ea
Therefore, using this parametrization,  we have
succeeded to disentangle the mass eigenvalues of $M$ from the diagonalizing matrix.
The eigenmasses $m_i$ are given  as functions of the coefficients  $b_i$
and the phase $\alpha$ only.  Consequently, for a given mass spectrum
we may reconstruct the fermion mass texture by simply computing the
coefficients $b_i$ from  relations (\ref{b1}-\ref{b3}) and a suitably chosen
unitary matrix $U$. If for example, the mixing effects are accurately described
by the experimental data, the mixing angles can be specified and $U$
 can be readily determined from the mixing matrix and the phase $\alpha$.
Next, we concentrate on the unitary matrix $U$
 assuming the standard parametrization in terms of
three angles $\theta_{12},\theta_{23},\theta_{13}$ and a phase $\delta$.
We have
\begin{equation}
U=%
\begin{bmatrix}
c_{12}c_{13} & c_{13}s_{12} & e^{-i\delta}s_{13}\\
-c_{23}s_{12}-c_{12}s_{13}s_{23}e^{i\delta} & c_{12}c_{23}-e^{i\delta}s_{12}s_{13}s_{23}
& c_{13}s_{23}\\
-c_{12}c_{23}s_{13}e^{i\delta}+s_{12}s_{23} & -c_{23}s_{12}s_{13}e^{i\delta}%
-c_{12}s_{23} & c_{13}c_{23}%
\end{bmatrix}\label{Unit}
\quad\cdot
\end{equation}
 The
requirement to have one eigenvalue equal to one leads to the  constraint
\begin{equation}
\sin \delta\sin\theta_{12}\sin\theta_{13}\sin\theta_{23}=0 \cdot
\end{equation}
This condition is satisfied if one of the parameters vanishes
generating four distinct structures for $U:$
\begin{equation}
U_{1}=%
\begin{bmatrix}
c_{12}c_{13} & c_{13}s_{12} & s_{13}\\
-c_{23}s_{12}-c_{12}s_{13}s_{23} & c_{12}c_{23}-s_{12}s_{13}s_{23} &
c_{13}s_{23}\\
-c_{12}c_{23}s_{13}+s_{12}s_{23} & -c_{23}s_{12}s_{13}-c_{12}s_{23} &
c_{13}c_{23}%
\end{bmatrix}
~,\quad \delta=0\label{dzero}
\end{equation}%
\begin{equation}
U_{2}=%
\begin{bmatrix}
c_{13} & 0 & e^{-i\delta}s_{13}\\
-s_{13}s_{23}e^{i\delta} & c_{23} & c_{13}s_{23}\\
-c_{23}s_{13}e^{i\delta} & -s_{23} & c_{13}c_{23}%
\end{bmatrix}
,\quad\theta_{12}=0\label{t12zero}
\end{equation}%
\begin{equation}
U_{3}=%
\begin{bmatrix}
c_{12} & s_{12} & 0\\
-c_{23}s_{12} & c_{12}c_{23} & s_{23}\\
s_{12}s_{23} & -c_{12}s_{23} & c_{23}%
\end{bmatrix}
,\quad\theta_{13}=0\label{t13zero}
\end{equation}%
\begin{equation}
U_{4}=%
\begin{bmatrix}
c_{12}c_{13} & c_{13}s_{12} & e^{-i\delta}s_{13}\\
-s_{12} & c_{12} & 0\\
-c_{12}s_{13}e^{i\delta} & -s_{12}s_{13}e^{i\delta} & c_{13}%
\end{bmatrix}
,\quad\theta_{23}=0\label{t23=0} \cdot
\end{equation}
 In the present analysis, out of the four possible forms for $U$ we choose the most
 appropriate cases and work out its implications on the leptonic sector. Since the mixing
effects depend on the combined charged lepton and neutrino diagonalizing matrices,
we treat them separately.

\subsection{The Neutrinos}

\bigskip

We start with the neutrino sector.  Here we do not assume a specific embedding  of the
matrices in a  given (GUT) model, thus the absolute scale of the neutrino eigenmasses is arbitrary
and can be easily chosen in consistency with the $\beta\beta$-decay constraints.

We will analyze the case where the
unitary matrix associated to the neutrino mass texture $M_{\nu}$ is expressed
as in (\ref{ExMU}) with $U$ given by the specific form (\ref{t12zero}), i.e.
\begin{equation}
U=%
\begin{bmatrix}
c_{13} & 0 & e^{-i\delta}s_{13}\\
-s_{13}s_{23}e^{i\delta} & c_{23} & c_{13}s_{23}\\
-c_{23}s_{13}e^{i\delta} & -s_{23} & c_{13}c_{23}%
\end{bmatrix}\cdot
\label{Uni4}%
\end{equation}
By construction, the above matrix admits one eigenvalue equal to one, thus
the eigenvalues of $U$ are $1$ and $e^{\pm i\alpha}$. The diagonalizing matrix
for $U$ is difficult to find in simple form, so we introduce the following
parametrization.
\begin{align}
\tan\theta_{13} &  =\frac{2z_{1}}{1-z_{1}^{2}}\\
\tan\theta_{23} &  =\frac{2z_{1}z_{2}\sqrt{\left(  1+z_{1}^{2}\right)  \left(
1-z_{2}^{2}\right)  }}{z_{2}^{2}-z_{1}^{2}+2z_{1}^{2}z_{2}^{2}}\\
\delta &  =\theta+\frac{\pi}{2}\cdot
\end{align}
Then, the diagonalizing matrix for $U$  is
\ba
V_{\nu}\left(  z_{1},z_{2},\theta\right)  =\frac{1}{\sqrt{2}}
\begin{bmatrix}
e^{i\theta}\frac{1}{p}\left(  z_{2}-\imath q^{2}\,z_{1}\right)   & \sqrt
{2}q\imath e^{i\theta} & -e^{i\theta}\frac{1}{p}\left(  z_{2}+\imath
q^{2}\,z_{1}\right)  \\
\frac{q}{p}\left(  z_{1}z_{2}-\imath\right)   & -\sqrt{2}z_{2} & \frac{q}
{p}\left(  z_{1}z_{2}+\imath\right)  \\
p & \sqrt{2}qz_{1} & p
\end{bmatrix}
\label{Vnu}
\ea
where $p,q$ are functions of $z_{1,2}$ given by
\begin{align}
p &  =\sqrt{\frac{1+z_{1}^{2}z_{2}^{2}}{\left(  1+z_{1}^{2}\right)  }%
}\label{Ap}\\
q &  =\sqrt{\frac{1-z_{2}^{2}}{1+z_{1}^{2}}}\ \cdot\label{arat}%
\end{align}
We can easily check that
\begin{equation}
V_{\nu}^{\dagger}UV_{\nu}=\mathrm{diagonal}\left[  e^{i\alpha},1,e^{-i\alpha
}\right]  \cdot
\end{equation}
The eigenvalue $\alpha$ depends only on  $z_{1},z_{2}$
\[
e^{i\alpha}=-\frac{z_{1}-iz_{2}}{z_{1}+iz_{2}},\;\mathrm{or}\;\alpha=\tan
^{-1}\frac{z_{1}}{z_{2}}\cdot
\]
The following relations are also useful:
\ba
z_{1}=\tan\frac{\theta_{13}}{2},&z_{2}=\frac{z_{1}\cos\frac{\theta_{23}}{2}
}{\sqrt{z_{1}^{2}+\sin^{2}\frac{\theta_{23}}{2}}}\nn
\\
p   =\sqrt{1-z_{2}^{2}\tan^{2}\frac{\theta_{23}}{2}},&
q =\frac{z_{1}}{z_{2}}\cot\frac{\theta_{23}}{2}\cdot\nn
\ea
Using these equations, all the elements of (\ref{Vnu}) can be expressed in
terms of the trigonometric entries of the unitary matrix $U$.

\bigskip

\subsection{The Charged Leptons}

\bigskip

We now derive  similar formulae for the charged leptons. As in the case of
neutrinos, we choose to write the mass matrix in terms of a unitary matrix $U$
in accordance to formula (\ref{ExMU}). For reasons that will become clear
later, we choose the ordering of the $U$ matrix eigenvalues to be as follows:
\begin{equation}
U=%
\begin{bmatrix}
1 & 0 & 0\\
0 & e^{i\alpha} & 0\\
0 & 0 & e^{-i\alpha}%
\end{bmatrix}
\ \cdot
\end{equation}
If the eigenvalues of $M$ are $m_{1},m_{2},m_{3}$ the coefficients $b_{i}$ in
(\ref{ExMU}) are given by equations analogous to (\ref{b1},\ref{b2},\ref{b3}).

For the case of charged leptons, we confine ourselves to orthogonal matrices.
Therefore, out of the four possible forms for $U$ we choose (\ref{dzero})
\begin{equation}
U=%
\begin{bmatrix}
c_{12}c_{13} & c_{13}s_{12} & s_{13}\\
-c_{23}s_{12}-c_{12}s_{13}s_{23} & c_{12}c_{23}-s_{12}s_{13}s_{23} &
c_{13}s_{23}\\
-c_{12}c_{23}s_{13}+s_{12}s_{23} & -c_{23}s_{12}s_{13}-c_{12}s_{23} &
c_{13}c_{23}%
\end{bmatrix}
\label{26}
\end{equation}
which corresponds to the structure $U_{1}.$  Next,  we use the fact that an
orthogonal matrix can be written as~\footnote{For a detailed mathematical
analysis of the subsequent formalism in the context of the fermion mass
matrices see~\cite{Leontaris:2009pi}.}
\[
U=e^{\alpha\widehat{n}\cdot\overrightarrow{s}}=1+\sin\alpha~\widehat{n}%
\cdot\overrightarrow{s}+\left(  1-\cos\alpha\right)  ~\left(  \widehat{n}%
\cdot\overrightarrow{s}\right)  ^{2}%
\]
where $\hat{n}=(n_{1},n_{2},n_{3})$ is a unit vector and the $3\times3$
matrices $s_{i}$ satisfy the conditions
\[
\left[  s_{i},s_{j}\right]  =\varepsilon_{ijk}s_{k}%
\]
and are explicitly given by
\begin{equation}
s_{1}=%
\begin{bmatrix}
0 & 0 & 0\\
0 & 0 & 1\\
0 & -1 & 0
\end{bmatrix}
\end{equation}%
\begin{equation}
s_{2}=%
\begin{bmatrix}
0 & 0 & 1\\
0 & 0 & 0\\
-1 & 0 & 0
\end{bmatrix}
\end{equation}%
\begin{equation}
s_{3}=%
\begin{bmatrix}
0 & 1 & 0\\
-1 & 0 & 0\\
0 & 0 & 0
\end{bmatrix}
\cdot
\end{equation}
The matrix $U$ is diagonalized by means of the matrix
\begin{equation}
V_{l}\left(  n_{1},n_{2},n_{3}\right)  =\frac{1}{\sqrt{2}\sqrt{n_{1}^{2}%
+n_{3}^{2}}}%
\begin{bmatrix}
\sqrt{2}\sqrt{n_{1}^{2}+n_{3}^{2}}n_{1} & n_{1}n_{2}-in_{3} & n_{1}%
n_{2}+in_{3}\\
-\sqrt{2}\sqrt{n_{1}^{2}+n_{3}^{2}}n_{2} & n_{1}^{2}+n_{3}^{2} & n_{1}%
^{2}+n_{3}^{2}\\
\sqrt{2}\sqrt{n_{1}^{2}+n_{3}^{2}}n_{3} & n_{2}n_{3}+in_{1} & n_{2}%
n_{3}-in_{1}%
\end{bmatrix}
\end{equation}
and we  may check that
\begin{equation}
V_{l}^{\dagger}UV_{l}=\mathrm{diagonal}\left[  1,e^{i\alpha},e^{-i\alpha
}\right]  \cdot
\end{equation}

\subsection{The Leptonic Mixing Matrix}

\bigskip
In the previous  sections we  managed to obtain the diagonalizing  matrices for the
charged lepton and neutrino mass textures $V_{l}$ and $V_{\nu}$ in closed form.
In accordance to standard notation,  the leptonic mixing matrix is defined to be
\ba
V_{M}=e^{i \psi}V_{l}^{\dagger}V_{\nu}\ \cdot\label{LepM}
\ea
The phase factor is introduced in order to have the matrix determinant equal to one.
A closer inspection reveals that all the elements of so derived $V_{M}$ are complex.
It can be shown~\cite{prepare} however,  that (\ref{LepM}) can be rendered equivalent
to the standard form given in (\ref{Unit}). The proof goes as follows.
A hermitean matrix $M$ can by diagonalized by means of a unitary transformation
\be
U^{\dagger }MU=\mathrm{Diag}\left[ m_{1},m_{2},m_{3}\right] =D\cdot \mathrm{Diag}
\left[ m_{1},m_{2},m_{3}\right] \cdot D^{\dagger }
\ee
where $D$ is a unitary matrix of the form
\be
D=\mathrm{Diag}\left[e^{ id_{1}},e^{ id_{2}},e^{ id_{3}}\right] \ \cdot
\ee
This way, if $U$ is a diagonalizing matrix so is $UD$ . The lepton mixing
matrix is given by
\begin{equation}
V_{M}=V_{l}^{\dagger}V_{\nu}\ \
\end{equation}%
and taking the above into account, $V_{M}$ can be equivalently written as
\be
V_{M}=D^{\dagger }V_{l}^{\dagger }V_{\nu }C\label{phas1}
\ee
where
\begin{eqnarray*}
C &=&\mathrm{Diag}\left[ e^{ ic_{1}},e^{ ic_{2}},e^{ ic_{3}}\right]  \\
D &=&\mathrm{Diag}\left[ e^{ id_{1}},e^{ id_{2}},e^{ id_{3}}\right] \ \cdot
\end{eqnarray*}%
If we require (\ref{phas1}) be reduced to the standard form, the six phases can be uniquely
determined.

\section{Analysis}

Using the above results, we will now proceed to determine
possible deviations from the TB-mixing which fit the experimental data
and determine the allowed range for the parameters $z_1,z_2,\theta,\hat n$ in
the neutrino and charged lepton sectors respectively. Before
 further pursuing the general case, we will first present the simplest
and possibly the most elegant way of extending the Tri-Bi maximal mixing.

\subsection{Example: The minimal case }

We start  with the neutrino sector and  introduce values for the parameters $z_{1,2}$
which are in accordance with the TB scenario. We  put $z_{2}=-1$ and we get $\tan\theta_{23}=0$
whilst for the eigenvalue of the unitary matrix $U$ we get
\[
e^{i\alpha}=\frac{i+z_{1}}{i-z_{1}}%
\]
so that $\tan\alpha=-\frac{2z_{1}}{1-z_{1}^{2}}$ and thus, $\alpha=-\theta_{13}.$
This way the $U$ matrix becomes
\begin{equation}
U=%
\begin{bmatrix}
\cos\alpha & 0 & -ie^{i\theta}\sin\alpha\\
0 & 1 & 0\\
-ie^{-i\theta}\sin\alpha & 0 & \cos\alpha
\end{bmatrix}\cdot
\end{equation}
The neutrino mass matrix takes the simple form
\ba
\begin{bmatrix}
\frac{1}{2}\left(  m_{1}+m_{3}\right)  & 0 & \frac{1}{2}e^{i\theta}\left(
m_{3}-m_{1}\right) \\
0 & m_{2} & 0\\
\frac{1}{2}e^{-i\theta}\left(  m_{3}-m_{1}\right)  & 0 & \frac{1}{2}\left(
m_{1}+m_{3}\right)
\end{bmatrix}
\label{TBmn} \cdot
\ea
For $\theta=\pi$ it is exactly the texture of the neutrino mass matrix
introduced in the case of TB mixing~\cite{Harrison:2002er}.
The matrix (\ref{TBmn}) is written in terms of its real eigenmass values  with
the $\{13\},\{31\}$ entries  multiplied by a phase factor.

Next, we proceed with the charged lepton sector.
The TB-matrix~\cite{Harrison:2002er} corresponds to
\begin{equation}
n_{1}=\frac{1}{\sqrt{3}}~,~n_{2}=-\frac{1}{\sqrt{3}}\ ,n_{3}=\frac{1}{\sqrt
{3}}\cdot
\end{equation}

We know however, that the TB case does not reproduce
 the experimental data since it  predicts a zero $\theta_{13}$ angle.
A minimal extension arises if we introduce in the neutrino mixing matrix the parameters
\[z_2=-1,\;\theta=\pi+\varphi\]
while  keeping the charged lepton diagonalizing matrix as above.
Then the mixing  matrix is given by

\be
V_{M}=e^{-\frac{i\pi}{6}}e^{-\frac{i\varphi}{3}} V_{l}^{\dag}\left(  \frac{1}{\sqrt{3}},-\frac{1}{\sqrt{3}},\frac
{1}{\sqrt{3}}\right) V_{\nu}\left(  z_{1},-1,\varphi+\pi\right) \label{Vmin} \cdot
\ee
After some  algebra and the removal of the redundant phase factors \cite{prepare},
 the matrix can be brought into canonical form given by
\be
V_{M}=%
\begin{bmatrix}
\sqrt{\frac{2}{3}}\cos\frac{\varphi}{2} & \frac{1}{\sqrt{3}} & -\sqrt{\frac{2}{3}%
}\sin\frac{\varphi}{2}\\
-\sqrt{\frac{2}{3}}\sin\left(  \frac{\varphi}{2}+\frac{\pi}{6}\right)   & \frac
{1}{\sqrt{3}} & -\sqrt{\frac{2}{3}}\cos\left(  \frac{\varphi}{2}+\frac{\pi}%
{6}\right)  \\
\sqrt{\frac{2}{3}}\sin\left(  \frac{\varphi}{2}-\frac{\pi}{6}\right)   & \frac
{1}{\sqrt{3}} & \sqrt{\frac{2}{3}}\cos\left(  \frac{\varphi}{2}-\frac{\pi}%
{6}\right)
\end{bmatrix}
\cdot
\ee
The experimental bounds are:
\begin{align}
0.0871557  & <\left\vert \sin\theta_{13}\right\vert <0.224931\\
0.68728  & <\left\vert \tan\theta_{12}\right\vert <0.713293\\
0.213895  & <\left\vert \tan\theta_{23}\right\vert <1.09131
\end{align}
and we have
\begin{align}
\left(  V_{M}\right)  _{11}  & =\sin\theta_{13}=-\sqrt{\frac{2}{3}}\sin\frac{\varphi}{2}\\
\frac{\left(  V_{M}\right)  _{23}}{\left(  V_{M}\right)  _{33}}  &
=\tan\theta_{23}=-\frac{\cos\left(  \frac{\varphi}{2}+\frac{\pi}{6}\right)  }{\cos\left(
\frac{\varphi}{2}-\frac{\pi}{6}\right)  }\\
\frac{\left(  V_{M}\right)  _{11}}{\left(  V_{M}\right)  _{12}}  &
=\tan\theta_{12}=\frac{1}{\sqrt{2}\cos\frac{\varphi}{2}}\cdot
\end{align}
Combining the above, we find that all the constraints are satisfied for
\begin{equation}
\frac{\pi}{15}\,\lesssim \,\varphi\,\lesssim \,\frac{\pi}{12}\label{tshift}\cdot
\end{equation}
This is a rather interesting result: it states that TB-mixing can reconcile the neutrino
data by a suitable choice of the phase parameter $\theta$ parametrizing the neutrino
diagonalizing matrix. In the minimal case we are here dealing with,
this phase coincides with the phase in the $\{13\},\{31\}$ elements
of the neutrino mass matrix. In the original TB model this phase is simply taken
to be $\theta=\pi$. When shifted by a
value $\varphi$  lying in the range (\ref{tshift}), neutrino data are exactly predicted.

The advantage of this solution as compared to any perturbative approach around the TB-solution
aiming to fit a nonzero $\theta_{13}$ angle is rather obvious: indeed, it is shown that a nonzero $\theta_{13}$  angle that preserves the symmetric and zero form texture of the $M_{\ell}$
and $M_{\nu}$ matrices can be naturally incorporated into the minimal TB-scheme.
 This is of crucial importance if we really wish to attribute their simple structure to
 some kind of discrete or other symmetry of the theory~\cite{Lam:2008sh}-\cite{deMedeirosVarzielas:2006fc}.

\subsection{The general case}

We explore now regions of the parameter space which signal departures from
the TB-case. Deviations  can be easily obtained by assuming for example that
\ba
n_{1}=\frac{1}{\sqrt{3}},\;n_{3}=\frac{1}{\sqrt{3}}-\varepsilon,\;n_{2}%
=-\sqrt{1-n_{1}^{2}-n_{3}^{2}}\label{Ved}
\ea
in the charged leptons sector. Similarly, we choose to write $z_{2}=-1+g^{2}$
 in the neutrino  diagonalizing matrix (\ref{Vnu}). In
figure we plot the ranges of these parameters subject to the well known
constraints of the mixing angles, while we keep $\theta=\pi$.

We observe that the allowed values  of $g$  lie in the range $0.05\lesssim$
$g\lesssim0.6$ , those of $\varepsilon$ in the range $-0.2\lesssim
\varepsilon\lesssim1$ while  acceptable values for $z_{1}$ cover a wider range
lying from $z_{1}\sim2.4$ to large negative values $z_{1}\sim-10$. It appears
that there are wide regions in the parameter space consistent with data which
significantly  deviate from the TB mixing picture.

\begin{figure}[t]
\centering \includegraphics [scale=0.8]{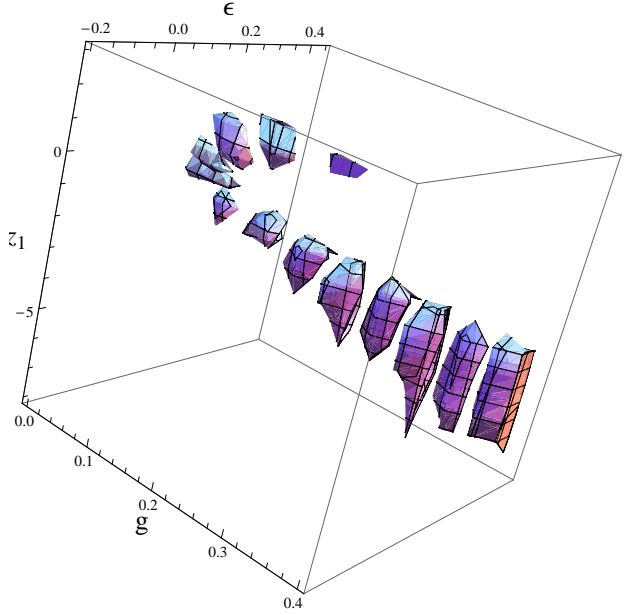}\caption{In this plot,
deviations from TB-mixing are parametrized in terms of $z_{1}$, $g^{2}%
=1+z_{2}$ of the neutrino and $\epsilon$ of the corresponding charged lepton
diagonalizing matrices. (see (\ref{Vnu},\ref{Ved}).)}%
\label{3Drange}%
\end{figure}

Next, in order to determine mass textures related to possible exact
symmetries, we scan the $g,\varepsilon$ ranges for  fixed $z_{1}$ values.
Here, we will concentrate in the subregions $g\sim\lbrack0.2-0.6]$,
$\varepsilon\sim\lbrack0.1-0.6]$ and search for values corresponding to
exactly known trigonometric quantities.

Let us choose $\varepsilon=1/\sqrt{3}$. This eliminates one entry in $V_{l}$
which assumes the form
\[
V_{\ell}=\left(
\begin{array}
[c]{ccc}%
\frac{1}{\sqrt{3}} & -\frac{1}{\sqrt{3}} & -\frac{1}{\sqrt{3}}\\
\sqrt{\frac{2}{3}} & \frac{1}{\sqrt{6}} & \frac{1}{\sqrt{6}}\\
0 & \frac{i}{\sqrt{2}} & -\frac{i}{\sqrt{2}}%
\end{array}
\right)  \ \cdot
\]
Upon inspection, we observe that the remaining two parameters of the neutrino
diagonalizing matrix can be taken to be $z_{1}=\tan\frac{7\pi}{12}$ ,
$z_{2}=-\frac{1}{\sqrt{2}}$ .

With this choice, we can readily check that the matrix formed by the moduli of
the elements of the leptonic mixing matrix is given by
\[
\left(
\begin{array}
[c]{ccc}%
0.806056 & 0.586939 & 0.0759986\\
0.420639 & 0.655601 & 0.627096\\
0.416337 & 0.475067 & 0.775225
\end{array}
\right)
\]
pretty much close to the experimental data shown below 
\[%
\begin{bmatrix}
\cdots& 0.546431-0.580416 & 0.03141-0.14091\\
\cdots& \cdots & 0.63505-0.736914\\
\cdots& \cdots & \cdots
\end{bmatrix}
\ \cdot
\]
where the missing elements are determined by unitarity.
For the above choice of parameters, the charged lepton mass matrix
obtained by substituting  (\ref{26}) into the general form (\ref{ExMU}), 
is found to be
\ba
M_{l}&=&\left(
\begin{array}
[c]{ccc}%
\frac{m_{1}+m_{2}+m_{3}}{3} & -\frac{2m_{1}-m_{2}-m_{3}}{3\sqrt{2}} &
-i\frac{m_{2}-m_{3}}{\sqrt{6}}\\
-\frac{2m_{1}-m_{2}-m_{3}}{3\sqrt{2}} & -\frac{4m_{1}+m_{2}+m_{3}}{6} &
-i\frac{m_{2}-m_{3}}{2\sqrt{3}}\\
i\frac{m_{2}-m_{3}}{\sqrt{6}} & i\frac{m_{2}-m_{3}}{2\sqrt{3}} & \frac
{m_{2}+m_{3}}{2}%
\end{array}
\right) \cdot\label{Mlgen}
\ea
The matrix  (\ref{Mlgen}) exhibits an interesting structure and one could think of
 several ways to link it to a possible existence of   underlying  symmetries.
 For example, in cases of  string derived models with several singlet fields $\phi,\phi',\dots$
 acquiring  vevs, one defines expansion parameters $\eps=\langle\phi\rangle/M$,
 $\eps'=\langle\phi'\rangle/M,\dots$ with $M$ being the cutoff scale of the higher theory.
Then, to leading order we get the hierarchy $m_3\gg m_2\gg m_1$ and we can approximate 
this matrix by
 \ba
M_{l}&\approx&\left(
\begin{array}
[c]{ccc}%
|\eps|^2 & \eps\bar\eps' &
\eps\\
\bar\eps\eps' & -{|\eps'|}^2 &
\eps'\\
\bar\eps & \bar\eps' & 1
\end{array}
\right) m_{\ell}\label{Mlapp}
\ea
with $\eps =i\sqrt{\frac{2}{3}},\eps'=\frac{i}{\sqrt{3}}$ and $m_{\ell}$ a mass parameter related
to charged lepton mass scale.

The corresponding neutrino mass matrix is given by
\[
M_{\nu}=\left(
\begin{array}
[c]{ccc}%
\frac{\alpha_+(m_{1}+\xi_-m_{2}+m_{3})}{4} & \frac{\beta_+(m_{1}-m_{3})+i\gamma_-(m_1-2m_{2}+m_{3})}{4} &
\frac{4\sqrt{2}(m_{1}-m_{3})-i(m_1-2m_{2}+m_{3})}{16}\\
\cdots& \frac{m_{1}+2m_{2}+m_{3}}{4} &
 \frac{i\beta_-(m_{1}-m_{3})+\gamma_+(m_1-2m_{2}+m_{3})}{4}\\
\cdots &\cdots & \frac{\alpha_-(m_{1}+\xi_+m_{2}+m_{3})}{4}
\end{array}
\right)
\]
where   the dots stand for the  corresponding complex conjugate entries and the various
coefficients are
\[
\alpha_{\pm}=\frac{6\pm\sqrt{3}}{4},\;\;\beta_{\pm}=
\frac{1\pm\sqrt{3}}{2},\;\;\gamma_{\pm}=\frac 12\sqrt{\tan\left({\pi}/4\pm{\pi}/6\right)}=\frac{\sqrt{2\pm\sqrt{3}}}{2},\;\;
\xi_{\pm}=\frac{32}{33}\alpha_{\pm}\gamma_{\pm}^2 \cdot
\]
It is to be noted that all the off-diagonal entries of the neutrino mass matrix are
expressed only in terms of the squared neutrino mass differences (note that
we have assumed Hermitian squared mass matrices thus we have the correspondence
$m_i\leftrightarrow m_{\nu_i}^2$).  The resulting structure is now  more complicated
than the corresponding  charged lepton one.
This is of course to be anticipated in models employing the see-saw  mechanism, since
the effective neutrino mass matrix is a product  of the Dirac
 and the heavy right handed Majorana neutrino mass matrices $M_{\nu}\propto m_DM_N^{-1}m_D^T$.
 Depending on the specific structure of the hypothetical  original theory, there are even
 more options to attribute this matrix to symmetry properties~\cite{Dreiner:1994ra},
 the analysis of this issue however goes beyond the scope of this letter.

\section{Conclusions}

In this letter we have investigated possible forms for the charged lepton and neutrino
mass textures  which can reconcile the experimental data on neutrino oscillations. In our
analysis we have considered the hermitian squares of either mass matrix and used standard
techniques to express each one of them as a second degree polynomial of a suitably chosen unitary
matrix. Since the eigenmass dependence is encapsulated  in the coefficients of this expansion
only, we can express the neutrino  mixing angles analytically, as functions of the parameters
which define the  unitary matrices that generate the charged lepton and neutrino mass textures
respectively.  Next, we may use the available neutrino data on
the mixing angles to constrain this parameter space. In particular, taking into account that
the mass matrices suggested by Harrison et al. are in good agreement with the Tri-Bi maximal
neutrino mixing, we explored the parameter space for allowed deviations. We have found
that the actual data including a non-vanishing $\theta_{13}$ angle can be nicely captured,
by only introducing a single phase in the $\{13\}$ and $\{31\}$ entries of the neutrino mass
 texture in the original TB-scheme. Furthermore, upon varying  the free parameters of our
 model in a wider range, we have found that neutrino data can be accommodated  even for large
 deviations from the TB-matrices too.

\end{document}